\documentclass[twocolumn,letter]{jpsj2} 
\usepackage{color}

\def\nle{\ \raise.3ex\hbox{$<$}\kern-0.8em\lower.7ex\hbox{$\sim$}\ }
\def\nge{\ \raise.3ex\hbox{$>$}\kern-0.8em\lower.7ex\hbox{$\sim$}\ }
\def\lcrh{L_h}
\def\Meq{M_{\rm eq}}

\def\Tc{T_{\rm c}}

\def\Tirr{T_{\rm irr}}

\def\Thlt{T_{\rm hlt}}

\def\FeMnTiOII{Fe$_{0.55}$Mn$_{0.45}$TiO$_3$}

\def\runauthor{Hajime \textsc{Takayama} and Koji \textsc{Hukushima} }

\title{Computational Experiment on Glassy Dynamic Nature of the
Field-Cooled Magnetization in an Ising Spin-Glass Model}
\author{ Hajime \textsc{Takayama}$^{1}$\footnote{E-mail address: takayama@issp.u-tokyo.ac.jp}   
and Koji \textsc{Hukushima}$^{2}$\footnote{E-mail address: hukusima@phys.c.u-tokyo.ac.jp} } 
\inst{$^{1}$Institute for Solid State Physics, University of Tokyo,
5-1-5 Kashiwa-no-ha, Kashiwa, Chiba 277-8581 \\
$^{2}$ Department of Basic Science, Graduate School of Arts and
Sciences, University of Tokyo, 3-8-1 Komaba, Tokyo 153-8902}
\recdate{\today}
\abst{The field-cooled (FC) process on an Ising spin-glass model is investigated by a standard Monte Calro (MC) method on one hand, and the equilibrium magnetization of the same system is evaluated by the exchang MC method on the other hand. 
The two types of simulation reveal intriguing glassy (nonequilibrium) dynamic properties of the FC magnetization (FCM) of the system.
Particularly, the FCM decrease is observed when the FC process is halted at a low temperater, although its value is smaller than the corresponding equilibrium value. 
We present a comprehensive interpretation of such peculiar phenomena based on the scenario for the FCM process recently propsed by J\"onsson and one of us (HT).
} 

\kword{Ising spin glass, field-cooled magnetization, droplet picture,
glassy dynamics}

\begin{document}
\sloppy
\maketitle

In the spin-glass (SG) study, one of most fundamental problems yet unsettled concerns the stability of the equilibrium SG phase under a finite magnetic field $h$~\cite{review1}.  
The SG mean-field theory predicts its stability up to a certain critical magnitude of $h$~\cite{AT-line}, while by the droplet theory~\cite{BM-chaos,FH-droplet} it is unstable even in an infinitesimal $h$.
There have appeared two representative experiments on an Ising spin glass in relatively large fields; one~\cite{KatoriIto94} favors the mean-field picture and the other~\cite{matetal95} does the droplet picture.  
Recently there have appeared numerical analyses on the Ising Edwards-Anderson (EA) model which strongly support the droplet picture.~\cite{HT-Huku,Young-Katz}
There still remains, however, a most fundamental SG property to be appropriately understood, namely, the behavior of the field-cooled magnetization (FCM) under a relatively small field. 

Since the early stage of the SG study, a kink-like behavior in the temperature dependence of FCM has been observed,~\cite{Nagata} and it has been considered to be evidence for the SG phase transition in a finite field.  
In detailed measurement on a AgFe spin glass, however, Lundgren {\it et al}~\cite{Lundgren} found the following peculiar FCM characteristics in small $h$.
The FCM  exhibits not a cusp but a peak at a temperature, denoted here as $T^*$, which is close to $\Tc$, a phase transition temperature under $h=0$.
Furthermore, when the FC process is halted at a temperature, denoted by $T_{\rm hlt}$, lower than $T^*$, the magnetization which we call the halted FCM (HFCM) initially decreases. 
By later experiments, the HFCM is checked to increase when $\Tirr < T_{\rm hlt} < T^*$ with $\Tirr$ being the so-called irreversibility temperature where the zero-field-cooled magnetization (ZFCM) starts to deviate from the FCM.~\cite{Nordblad}
The changes in FCM involved in these phenomena are small in magnitude as compared with a value of the FCM itself, but they are considered to be intrinsic properties commonly shared by most of typical spin glasses.
Having further confirmed these intriguing FCM properties by both experiments on an Ising spin glass \FeMnTiOII\ and by numerical simulations on the Ising EA model, J\"onsson and Takayama (JT)~\cite{Petra-HT} have proposed, based on the SG droplet picture, a novel scenario which explains these FCM properties as the consequence of the glassy (non-equilibrium) nature peculiar to the FCM dynamics involved.~\cite{Petra-icm}

In the JT analysis, however, important information has been missing, i.e., the one on the equilibrium magnetization (EQM) for a given set of temperature and field, $\Meq(T,h)$, which is the destination of the HFCM at the corresponding ($T, h$). 
To our knowledge, there has been no real experiment, in which the HFCM is observed to reach the EQM.
This is consistent with the intrinsic SG nature, i.e., the marginal stability predicted by both mean-field and droplet theories. 
It implies that relaxation times are distributed continuously up to infinity not only at the transition point but also in a whole SG phase.
By computational experiments, on the other hand, we can evaluate the EQM of a finite system by using the sophisticated numerical methods such as the exchange Monte Carlo (MC) method~\cite{Huku-Nemoto} which artificially accelerate equilibration in SG systems.
We can, of course, analyze the FCM of the same system by a standard MC method.  
The information on the EQM is quite important particularly for analyzing dynamical processes in such a system with the marginal stability. 
In fact, the main result of the present work is that the HFCM at $T_{\rm hlt} < T^*$ does decrease though it is smaller than the corresponding EQM, indicating that the phenomenon is far from equilibrium.
The purpose of the present Letter is to report the FCM behavior simulated further in details as well as the EQM data, and then to argue comprehensively such intriguing FCM phenomena as the FCM peak, the HFCM decrease at $T_{\rm hlt} < T^*$, and its increase at $T^* < \Thlt < \Tirr$, based on the JT scenario.

\begin{figure}[t]
\begin{center}
\resizebox{0.5\textwidth}{!}{\includegraphics{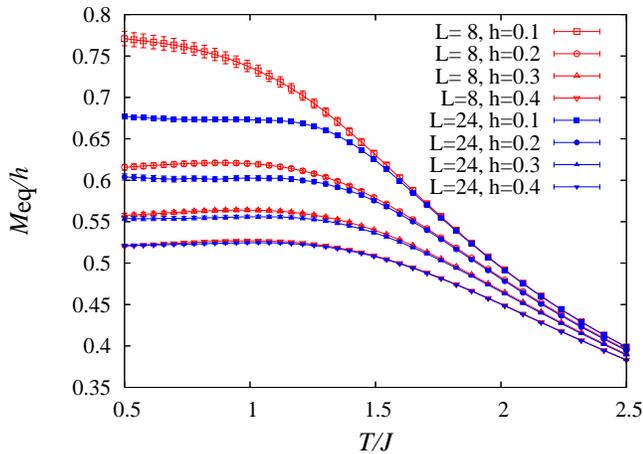}}
\end{center}
\caption{$\Meq (T;,h,L)$ evaluated by the exchange MC method.}
\label{fig:Meq}
\end{figure}

The system we examine is the 3-dimensional Ising EA model whose nearest neighbor interactions obey a Gaussian distribution with mean zero and variance $J$ which we use as the units of energy and temperature (with $k_{\rm B}=1$). 
Its $\Tc$ has recently been estimated as $\Tc \simeq 0.96$.~\cite{Katz-K-Y}
The strength of field $h$ is represented by its associated Zeeman energy. 
We adopt a standard heat-bath MC method to simulate FC processes, and the time is measured in unit of MC steps per spin (mcs).
A system is cooled by a step of $\Delta T=0.01$, while it is held by a staying time, $t_{\rm sty} = n$ mcs, at each temperature. 
We call this a rate $n$, or r$n$, cooling process. 
To specify a value of the FCM per spin, we take an average over at most last 30 mcs at each $T$, which is much shorter than $n$. 
For the FCM analysis, systems only with the linear dimension $L=24$ are simulated as in our previous works.~\cite{Petra-HT,ours1} 
We take averages over several thousands or more samples to get sufficient accuracy for discussion of small differences in FCM and HFCM of interest. 
To evaluate the EQM we carry out the exchange MC simulation~\cite{Huku-Nemoto} with 64 temperature points in the range $0.5 \le T \le 2.75$, and take average over 1500 samples.
The periodic boundary condition is imposed for systems simulated by both methods. 

Let us begin our discussion with the EQM results shown in Fig.~\ref{fig:Meq}. 
The EQM's of the systems with $L=8$ and 24 coincide with each other even at lower temperatures than $\Tc$ when $h$ is large ($=0.4$), but they don't for small $h$ such as $h=0.1$. 
It implies the presence of a certain characteristic length scale which is smaller (larger) than $L=8$ under $h=0.4\ (0.1)$. 
In the SG droplet theory, we immediately think of such a length scale, namely, the field crossover length, $\lcrh$. 
It separates the characteristic behavior of droplet excitations in the equilibrium SG state by their size $l$; it is dominated by the Zeeman energy ($\sim hl^{d/2}$) for $l>\lcrh$ and by the SG stiffness energy ($\sim \Upsilon \sqrt{q_{\rm EA}}l^\theta$) for $l<\lcrh$. 
Here, $d$ is the spatial dimension, $\Upsilon$ the stiffness constant of the SG ordering, $q_{\rm EA}$ the EA order parameter, and $\theta$ the stiffness exponent. 
Explicitly, $\lcrh$ is written as  
\begin{equation}
\lcrh = l_T h^{-\delta}, 
\label{eq:lcrh}
\end{equation}
where $\delta = (d/2 - \theta)^{-1}$ and $l_T$ is a constant weakly depending on $T$ through $q_{\rm EA}$ and $\Upsilon$. 
Then the $L$-independent EQM at sufficiently low temperatures simply implies that $\lcrh<L=8$ for a large field ($h=0.4$).

\begin{figure}[b]
\begin{center}
\resizebox{0.5\textwidth}{!}{\includegraphics{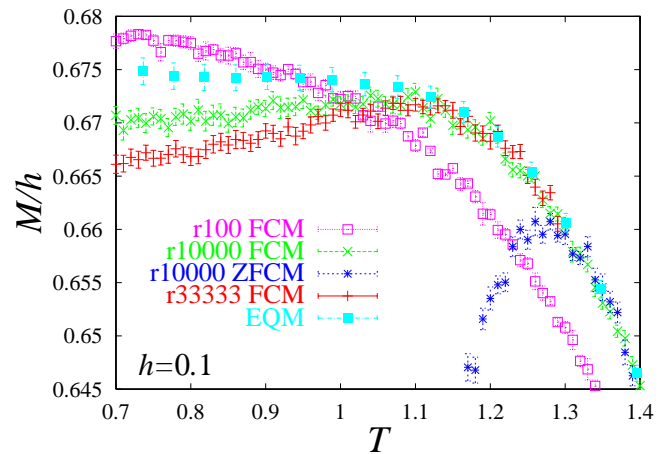}}
\end{center}
\caption{The FCM, ZFCM and EQM in $h=0.1$ ($L=24$). 
}
\label{fig:Mfc}
\end{figure}

The simulated $L=24$ EQM seen in Fig.~\ref{fig:Meq} is nearly constant up to a temperature much higher than $\Tc$.
This observation leads us to extend the presence of $\lcrh$ to a temperature region higher than $\Tc$ in the following sense.
As the temperature is decreased from above, the SG short-range-order (SRO) grows but is hindered by field $h$ around a temperature, denoted as $T_{\rm crs}$, where the SG coherence length, $\xi_{\rm c}(T)\ \propto (T-\Tc)^{-\nu}$, becomes comparable with the extended $\lcrh$. 
These SRO's grow inhomogeneously in space due to the randomness in degrees of frustration in the SG system, and they are considered to behave as randomly frozen clusters at $T \nle T_{\rm crs}$. 
Namely, each such cluster is fluctuating as a whole in equilibrium but with a reduced magnetization per spin and so the response of the system to $h$ is reduced.
The consequence is the EQM smoothed-kink behavior seen in Fig.~\ref{fig:Meq}, from which we roughly specify a value of $T_{\rm crs}$. 
The parameters $\Upsilon$ and $q_{\rm EA}$ involved in the extended $\lcrh$ are considered to be the corresponding quantities associated with the SG SRO. 
Without going into its details, we simply assume here that the temperature dependence of the extended $\lcrh$ is significantly weaker than that of $\xi_{\rm c}(T)$, and that the extended $\lcrh$ smoothly changes to the original one at low temperatures. 
It is also noted that the above argument is appropriate only for the case $L > \lcrh$. 
For systems with $L < \lcrh$, we have to take into account the finite-size effect introduced by the periodic boundary condition.   
The $L=8$ EQM in $h=0.1$, which deviates from the $L=24$ EQM around $T_{\rm crs}$, is the case. 
Its details will be discussed elsewhere.

Now let us move to discussion on the FCM behavior.
In Fig.~\ref{fig:Mfc} we show FCM's observed in three FC processes with different cooling rates as well as the EQM in $h=0.1$. 
Also shown are the ZFCM observed in the r10$^4$ process. 
One can see that $\Tirr \simeq 1.3$ for the r10$^4$ process, whereas it is out of the figure, $\Tirr \simeq 1.6$, for the r10$^2$ process (see Fig. 1 in [\citen{Petra-HT}]). 
The important observations here are that the FCM's with slow cooling (r10$^4$ and r33333) exhibit a peak, and that these FCM'S do not exceed the EQM even at $T \simeq T^*$. 
The peak is the sharper, the slower is the cooling.
In contrast to real experiments,~\cite{Lundgren, Petra-HT} the crossing of FCM's below $T^*$ is not ascertained in the present simulation.
The reason can be attributed to the field strength, namely $h=0.1$ is still large for the crossing to be observed.
The FCM with rapid cooling (r10$^2$), on the other hand, does not exhibit a peak, and it even exceeds the EQM at low temperatures. 

\begin{figure}[t]
\begin{center}
\resizebox{0.48\textwidth}{!}{\includegraphics{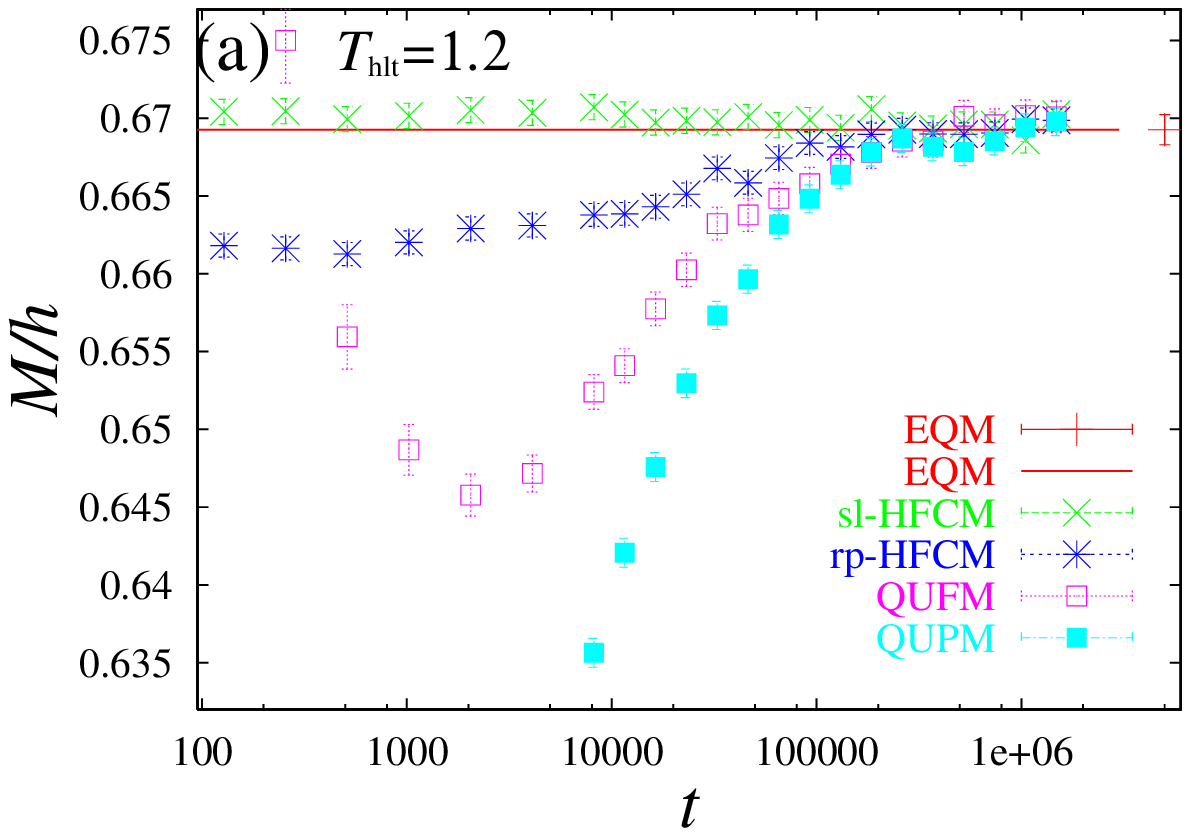}}
\resizebox{0.48\textwidth}{!}{\includegraphics{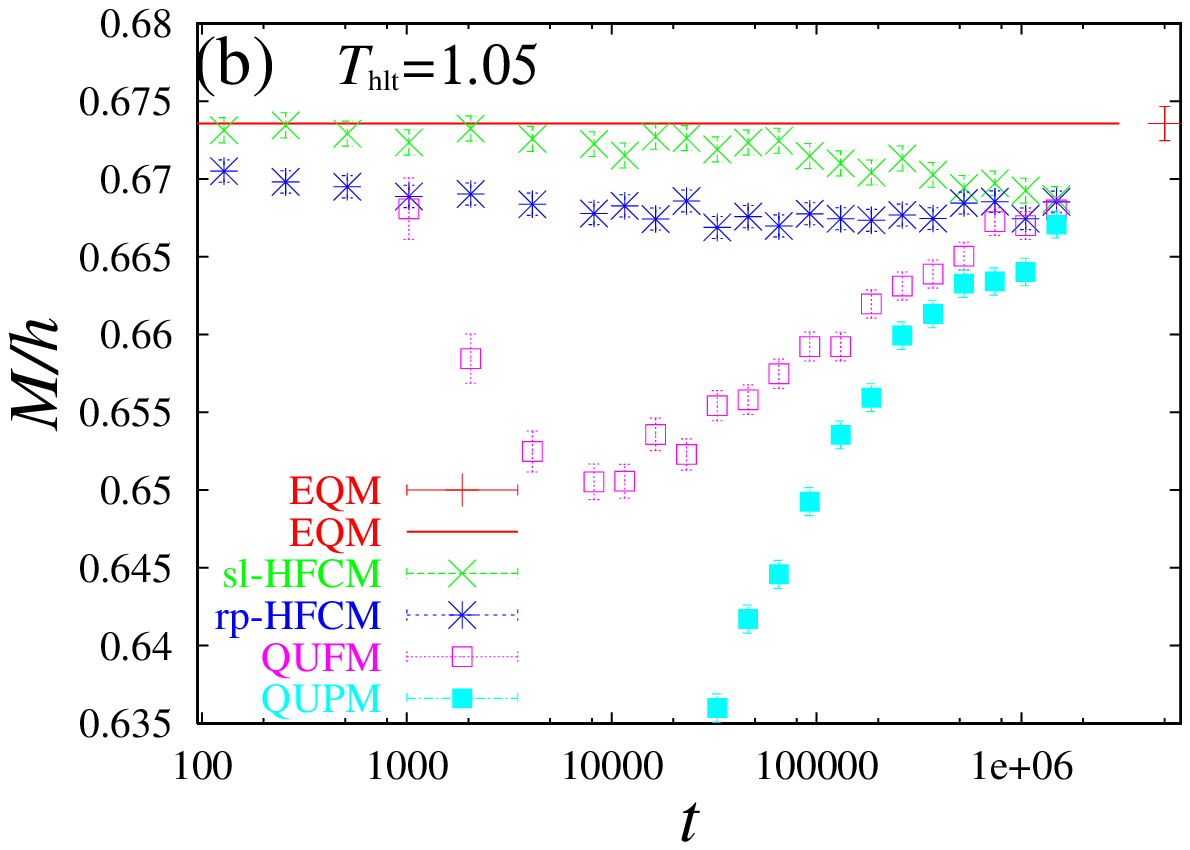}}
\resizebox{0.48\textwidth}{!}{\includegraphics{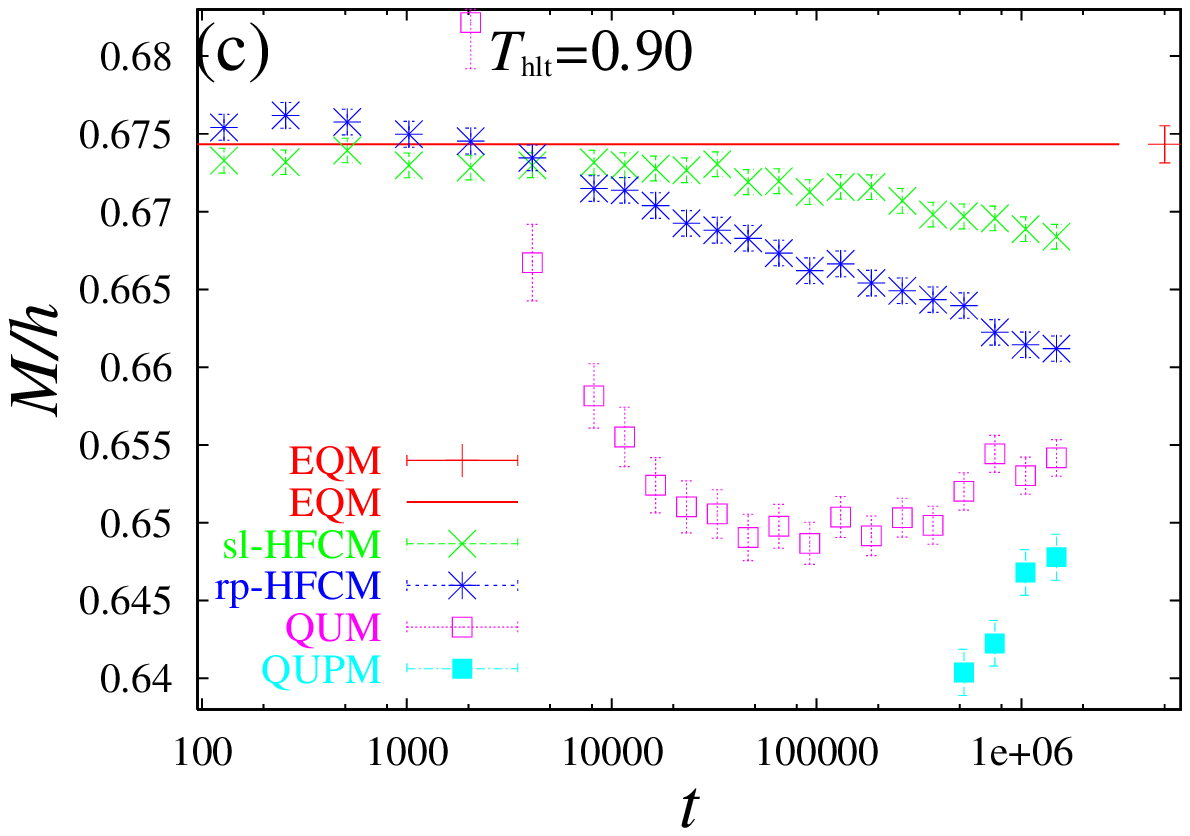}}
\end{center}
\caption{The HFCM, QUFM and QUPM in $h=0.1$ at three $T_{\rm hlt}$'s. The horizontal line represents the EQM with its error bar indicated at the right end.}
\label{fig:at-stop}
\end{figure}

The HFCM results at three $\Thlt$'s after the rapid (r10$^2$) and slow (r10$^4$) FC processes are shown in Fig.~\ref{fig:at-stop}. 
Also shown there are time evolutions of magnetizations after an instantaneous quench from the paramagnetic (completely random) and ferromagnetic (completely polarized) states at the same temperature, denoted respectively as QUPM and QUFM. 
We see in Fig.~\ref{fig:Mfc} that the equilibrium is realized by the slow cooling down to $T \nle 1.1\ (< \Tirr)$ within the present numerical accuracy . 
Correspondingly, its HFCM at $\Thlt=1.2$ stays constant as seen in Fig.~\ref{fig:at-stop}(a).
The FCM in the rapid cooling, on the other hand, is significantly smaller than the EQM already at $T=1.2$ and its HFCM at $\Thlt=1.2$ starts to increase at time $t \nge n \ (=10^2)$. 
Its equilibration is accomplished at $t \simeq 10^5$, where both QUPM and QUFM also reach to the EQM. 
Interestingly, the QUFM does so from below, after it overshoots the EQM.  
At a low temperature such as $T=0.9$, the FCM in the slow cooling, having passed its peak, is smaller than the EQM (Fig.~\ref{fig:Mfc}).
Still, as seen in Fig.~\ref{fig:at-stop}(c), its HFCM at $\Thlt=0.9$ definitely starts to decrease, which is the opposite direction to the EQM. 
Also the HFCM at $\Thlt=0.9$ after the rapid cooling starts to decrease from its initial value larger than the EQM, continues to decrease, and overshoots the EQM.
We can see also in the figure that the QUFM exhibits a minimum at $t \simeq 10^5$, but it is still significantly smaller than the two HFCM's even at $t \simeq 10^6$, where the QUPM is further smaller than the QUFM.
These data suggest that it will take $10^{10}$ mcs or much more for these magnetizations to reach the EQM at this temperature.
At $\Thlt=1.05$ shown in Fig.~\ref{fig:at-stop}(b), a little lower than $T^*$ of the slow FC process, the corresponding HFCM exhibits monotonous decrease. The HFCM after the rapid cooling, on the other hand, exhibits a subtle upturn after its initial decrease. 
These results strongly suggest that a certain qualitative change in nature of the FCM dynamics occurs around this temperature also in the rapid cooling, though its FCM does not exhibit a peak. 

Now let us interpret the peculiar FCM phenomena so far described along the line of the JT scenario.
It has been well accepted that, at $T \simeq \Tirr$, spin clusters of a mean size denoted as $\xi^*$, go into (release from) their thermally blocked state in the cooling (reheating) process of the FCM (ZFCM) measurement.
Here `thermally blocked' means that clusters are not fluctuating within time scale of $t_{\rm sty}$ of the FC and ZFC processes adopted.
Therefore, $\Tirr$ depends on both $h$ and $t_{\rm sty}$, while $T_{\rm crs}$ for the EQM does only on $h$.
Correspondingly, $\xi^*$ also depends on $h$ and $t_{\rm sty}$.
Furthermore, we consider it plausible that these thermally blocked clusters appear inhomogeneously in space and are isolated with each others.
As the FC process is continued below $\Tirr$, clusters whose sizes are smaller than $\xi^*$ become thermally blocked.
But when the FC process is halted by a much longer time than $t_{\rm sty}$, the clusters already thermally blocked become able to fluctuate and the HFCM increases toward the EQM. 

According to the JT scenario, the spin clusters so far thermally blocked and isolated from each others become in touch with at $T \simeq T^*$.
Then the SG stiffness energy, so far effective only within each clusters, now becomes effective also between the clusters, and the latter rearrange themselves to a configuration which gains the SG stiffness energy.  
This rearrangement is initiated by interactions with neighboring spins. 
It therefore starts to occur in a length scale much smaller than $\xi^*$, which in turn is smaller than $\lcrh$ in the FC process in small $h$ and with small $t_{\rm sty}$. 
Its consequence is, of course, decrease in the total magnetization of the system. 
Even at $T<T^*$, however, there may be still smaller clusters which are isolated from others and which are going to thermally blocked preferentially to the field direction by the temperature decrease of $\Delta T$. 
In the rapid FC process, the latter dominates the former, giving rise to only decrease of the FCM increasing rate but not the FCM peak. 
The opposite is the case in the slow FC process, yielding the FCM peak.

As the temperature is further decreased, cluster flips governed by the thermally activated process in this temperature range and below become less and less frequent and the FCM becomes nearly constant. 
But there certainly still exist isolated spins (and small clusters) whose internal field acting on them, denoted as $h_{\rm int}$, is vanishingly small. 
Actually, the distribution of $h_{\rm int}$'s of the mean-field model is shown to be an even function of $h_{\rm int}$ and is zero at $h_{\rm int}=0$ in the limit $h \rightarrow 0$.~\cite{Nemoto}
This is naturally expected to $h_{\rm int}$'s in the EA model of the present interest as well.
These isolated spins and small clusters are expected to be blocked equally to both directions relatively to the field on average, yielding the constant FCM.
Similar FCM behavior is observed in a superspin glass consisting of magnetic fine particles interacting with each others via dipole-dipole interactions.~\cite{Sasaki-etal}

When the FC process is halted at $T<T^*$, the rearrangement of clusters mentioned above proceeds and so the HFCM decreases. 
As seen in Fig.~\ref{fig:at-stop}(c), this happens even at a halt of the rapid FC process after an interval shorter than $t_{\rm sty}$ where the thermal blocking of small clusters due to the temperature decrease of $\Delta T$ have completed.
This HFCM decrease is expected to continue until the mean size of the SG SRO proper to $T=\Thlt$ and $h=0$, denoted as $\xi_T(t)$, reaches to $\lcrh$.
At further longer time, $\xi_T(t)$ no longer grows. 
Instead, the system consisting of clusters of a mean size $\lcrh$ relaxes to its equilibrium configuration, yielding the HFCM upturn and its increase toward the EQM.
The HFCM upturn after the initial decreases is vaguely seen in the rapid FC process as shown in Fig.~\ref{fig:at-stop}(b).
The QUFM upturn after the initial rapid decrease seen in Fig.~\ref{fig:at-stop} can be similarly interpreted.

The computational results so far described agree qualitatively with those observed in real experiments on various typical spin glasses, though the scales of time and field strength between the two types of observation quantitatively differ very much.
In our previous work on the field-shift aging protocol on Ising spin glasses,~\cite{HT-Huku} the computational result is extended to the laboratory time range by making use of the growth law of the above-mentioned $\xi_T(t)$ which is also obtained numerically.~\cite{ours1}
This reproduces even semi-quantitatively the corresponding experimental result,~\cite{KatoriIto94} implying that the two measurements look at the same physics.
We believe that this is also the case for the present FCM analysis in a relatively weak field.

Recently non-equilibrium phenomena below $T^*\ (\sim \Tc)$, such as aging, memory, and rejuvenation, have been extensively studied.~\cite{review1}  
Although most of the results are interpreted by regarding that the state reached by a cooling is in near equilibrium, that is in fact not the case as clearly demonstrated by the HFCM decrease at $\Thlt < T^*$ in the present work. 
We consider that quasi-equilibrium interpretation is appropriate only for phenomena within a time scale of the order of the waiting time adopted, such as $t_{\rm sty}$ in the present analysis.
If the measurement is continued to much longer time, the non-equilibrium nature discussed in the present work is expected to be observed.
In fact, the post-aging decay phenomenon in such a time range has been recently reported.~\cite{Kenning} 
We consider it to be due to the presence and very slow decay of clusters at $T<\Tc$ which have been thermally blocked at $T \simeq \Tirr > \Tc$, or more generally, at $T$ where the characteristic relaxation time of the SR SRO of a size $\xi_{\rm c}(T)$ exceeds $t_{\rm sty}$ of a cooling process adopted.

To conclude, by computational experiments, or by numerically looking at physical properties that a well-defined, rather simple theoretical (Ising EA SG) model exhibits,  we have observed its intriguing FCM phenomena such as the FCM peak and the HFCM increase (decrease) at a halt at $T^* < \Thlt < \Tirr$ ($\Thlt < T^*$).
The computational experiment can further evaluate equilibrium properties by artificially accelerating the relaxation dynamics.
Combined the EQM result thus obtained with the FCM behavior, we have presented a comprehensive interpretation of most of the properties associated with the FC process based on the JT scenario.
Although we have introduced certain assumptions such as the presence of the extended field crossover length up to temperatures above $\Tc$, we rather consider the latter one of the conclusions reached by the present computational experiment.

We thank to P. E. J\"onsson for valuable discussions.
The present work was supported by the Next Generation Supercomputing Project, Nanoscience Program and the Grants-In-Aid for Scientific Research (No.~17540348 and No.~18079004), both from MEXT of Japan.

\end{document}